\documentclass[a4paper,11pt]{article}
\usepackage{pos}
\usepackage{subcaption}
\usepackage{microtype}

\DeclareMathOperator{\tr}{tr}
\DeclareMathOperator{\re}{Re}
\DeclareMathOperator{\im}{Im}

\title{Neural Network Gauge Field Transformation for 4D SU(3) gauge fields}

\author*[a]{Xiao-Yong Jin}

\affiliation[a]{Computational Science Division, Argonne National Laboratory,\\
	Lemont, IL 60439, USA}

\emailAdd{xjin@anl.gov}

\abstract{%
We construct neural networks that work for any Lie group and maintain
gauge covariance, enabling smooth, invertible gauge field transformations. 
We implement these transformations for 4D SU(3) lattice gauge fields
and explore their use in HMC.
We focus on developing loss functions and optimizing the transformations.
We show the effects on HMC's molecular dynamics and discuss
the scalability of the approach.
}

\FullConference{The 40th International Symposium on Lattice Field Theory (Lattice 2023)\\
July 31st - August 4th, 2023\\
Fermi National Accelerator Laboratory\\}

\begin{document}
\maketitle

\section{\label{sec:intro}Introduction}

Neural networks can approximate
arbitrary functions~\cite{Cybenko:1989iql,Hornik:1989yye,Hornik:1991sec,LESHNO1993861}.
When combined with gauge symmetries~\cite{pmlr-v97-cohen19d,pmlr-v119-finzi20a}
and adapted to lattice fields~\cite{Luo:2020stn,Favoni:2020reg,Nagai:2021bhh,Jin:2022bgq,Aronsson:2023rli,Tomiya:2023jdy},
they offer new tools for developing algorithms~\cite{Kanwar:2020xzo,Boyda:2020hsi,Foreman:2021ixr,Boyda:2022nmh,Abbott:2022zhs,Lehner:2023bba,Lehner:2023prf,Nagai:2023fxt} in lattice QCD.
One key application is improving the efficiency of the Hybrid Monte Carlo (HMC) algorithm~\cite{Duane:1987de}.

HMC uses Hamiltonian dynamics
to generate proposal configurations for a Markov Chain,
sampling the configuration space according to a known unnormalized probability density function.
The Hamiltonian dynamics, however,
can struggle with potential barriers due to limited energy
or may linger in lengthy potential valleys.
Lattice QCD systems, with their high degrees of freedom and separation of long and short distance modes,
exhibit critical slowing down~\cite{Schaefer:2010hu}
near the continuum limit,
with the cost of generating independent gauge field configurations also increasing
as the force in the Hamiltonian dynamics requires finer numerical integration steps.
This conference offers two reviews~\cite{Kanwar:2024ujc,Boyle:2024nlh} on improvements to HMC algorithms.

This work focuses on creating smooth and invertible gauge field transformations with tunable parameters,
incorporating neural networks while ensuring gauge covariance.
Generalizing prior work on the perturbative Wilson flow~\cite{Luscher:2009eq}
and extending a study in 2D U(1) gauge field~\cite{Jin:2022bgq},
we apply tuned transformations in HMC with 4D SU(3) gauge fields.

We first describe the gauge field transformation,
then introduce the neural networks,
and present results in optimizing and applying the transformation in molecular dynamics.
We discuss the scalability and transferability of our approach across different lattice volumes and gauge couplings.

\section{\label{sec:transform}Gauge field transformation}

We introduce a continuously differentiable bijective map $\mathcal{F}$
with $U=\mathcal{F}(V)$ on the gauge fields $V \mapsto U$,
and apply a change of variables in the path integral for an observable
$\mathcal{O}$,
\begin{equation}
	\langle \mathcal{O}\rangle =
	\frac{1}{\mathcal{Z}} \int
	\mathrm{D}[U] \mathcal{O}(U) e^{-S(U)} =
	\frac{1}{\mathcal{Z}} \int
	\mathrm{D}[V] \mathcal{O}\big(\mathcal{F}(V)\big) e^{-S_{\text{FT}}(V)},
\end{equation}
with the effective action after the field transformation,
\begin{equation}
	\label{eq:effective-action}
	S_{\text{FT}}(V) =
	S\big(\mathcal{F}(V)\big) - \ln\big|\mathcal{F}_*(V)\big|,
\end{equation}
and the Jacobian of the transformation,
\begin{equation}
	\mathcal{F}_*(V) = \frac{\partial\mathcal{F}(V)}{\partial V}.
\end{equation}

We construct the gauge field transformation with
\begin{equation}
	\label{eq:transform}
	U_{x,\mu} = \exp\left[\sum_l \epsilon_{x,\mu,l} \partial_{x,\mu} W_{x,\mu,l}\right] V_{x,\mu},
\end{equation}
where $V_{x,\mu}$ denotes the gauge link from lattice site $x$ to $x+\hat{\mu}$,
in the direction $\mu$.
The Wilson loop, $W_{x,\mu,l}$,
is an ordered product of gauge links along the loop labeled $l$ that goes through the link at $x,\mu$.
The group derivative, $\partial_{x,\mu} W_{x,\mu,l}$, is
with respect to the gauge link, $V_{x,\mu}$.

Generalizing the stout smear~\cite{Morningstar:2003gk} or the Wilson flow~\cite{Luscher:2009eq},
we optionally make
the coefficients depend on Wilson loops ($X$, $Y$, \ldots) independent of $V_{x,\mu}$,
\begin{equation}
	\label{eq:def-eps}
	\epsilon_{x,\mu,l} = c_l \frac{2}{\pi} \tan^{-1}\big[ \mathcal{N}_l(X,Y,\ldots) \big].
\end{equation}
The additional application of $\tan^{-1}$ and the multiplication of scalar
coefficients $c_l$ serve to constrain the possible values of $\epsilon$,
ensuring the resulting Jacobian determinant is positive definite.
The independence of the Wilson loops ($X$, $Y$, \ldots) from $V_{x,\mu}$
simplifies the computation of the Jacobian determinant
and results in the constraint~\cite{Luscher:2009eq}
\begin{equation}
	c_l = \frac{3}{4 N_l},
\end{equation}
for $N_l$ the number of distinct Wilson loops $l$.
Naturally, $\mathcal{N}_l$ can be any neural network.

In practice, we update only a subset of the gauge links at a time,
a single direction out of four space-time directions,
and on either even or odd lattice sites lexicographically.
For each subset, we pick the derivatives of the Wilson loops,
$\partial_{x,\mu} W_l$,
such that the only link to be updated in $W_l$ is $x,\mu$.
This ensures that the Jacobian matrix of the transformation is diagonal.
Updating the whole lattice requires eight separate subset updates.

\begin{figure}
	\centering
	\includegraphics[width=0.618\textwidth]{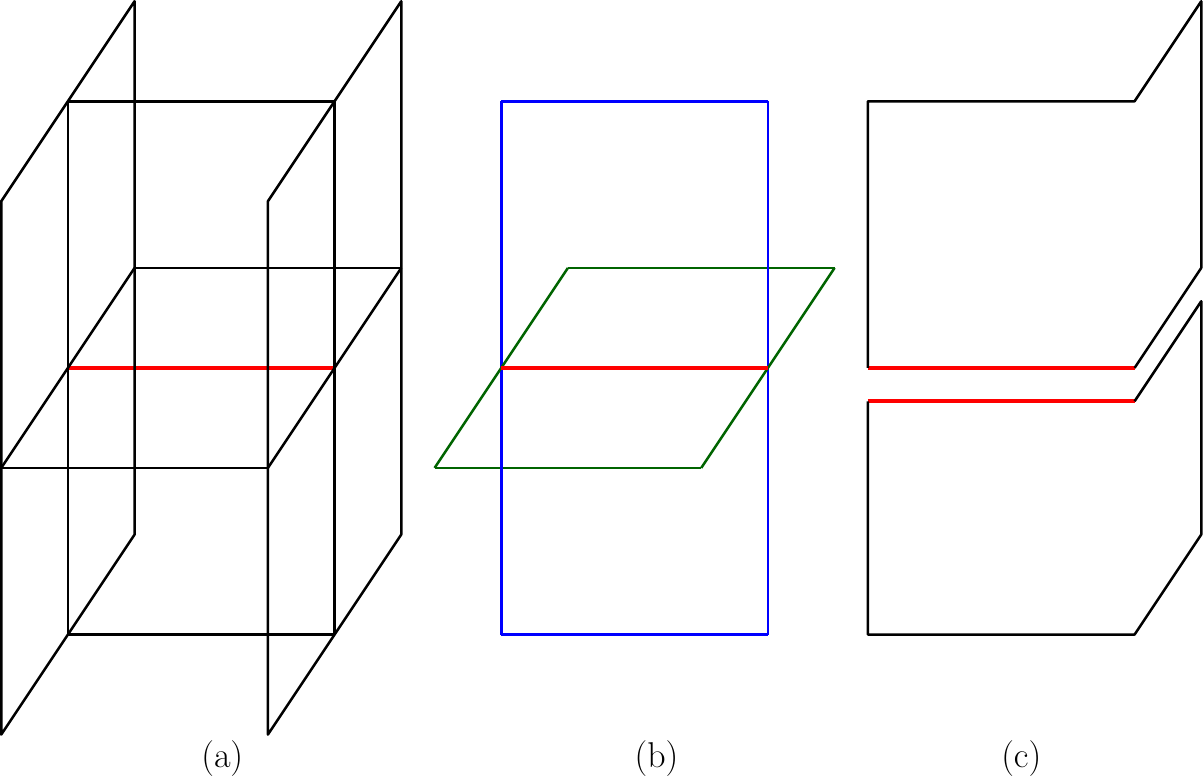}
	\caption{\label{fig:smearing}Wilson loops used in smearing update to the red link in a 3D lattice.
	From left to right, (a) the links in black used to update the red link, (b) four plaquette
	loops, (c) two 6-link chair shaped loops with one side perpendicular to, and at the right side of, the red link.}
\end{figure}

Figure~\ref{fig:smearing} shows the Wilson loops used for taking derivatives,
in equation~\eqref{eq:transform}, when updating a single gauge link, in our implementation.
While the figure depicts a 3D lattice, in 4D space-time,
there are 6 plaquette loops and 48 chair loops,
which go through the updating link marked red in the figure.
The derivatives of those loops with respect to the red link generate the basis,
the linear combination of which forms the $su(3)$ algebra space in the exponential map in equation~\eqref{eq:transform}.

\section{Neural network architectures}

Stacking layers of individual transformations as in equation~\eqref{eq:transform}
forms a deep residual neural network~\cite{Nagai:2021bhh} generalized for gauge fields.
Further generalizing it with standard neural networks,
we employ basic neural network layers that sequentially apply to traced Wilson loops
and produce localized smearing coefficients $\epsilon_{x,\mu,l}$ as introduced in equation~\eqref{eq:def-eps}.
As mentioned,
to simplify the computation of the Jacobian matrix,
the traced Wilson loops used at each layer are independent of the links to be updated at that layer with equation~\eqref{eq:transform}.

\begin{figure}
	\centering
	\includegraphics[width=0.618\textwidth]{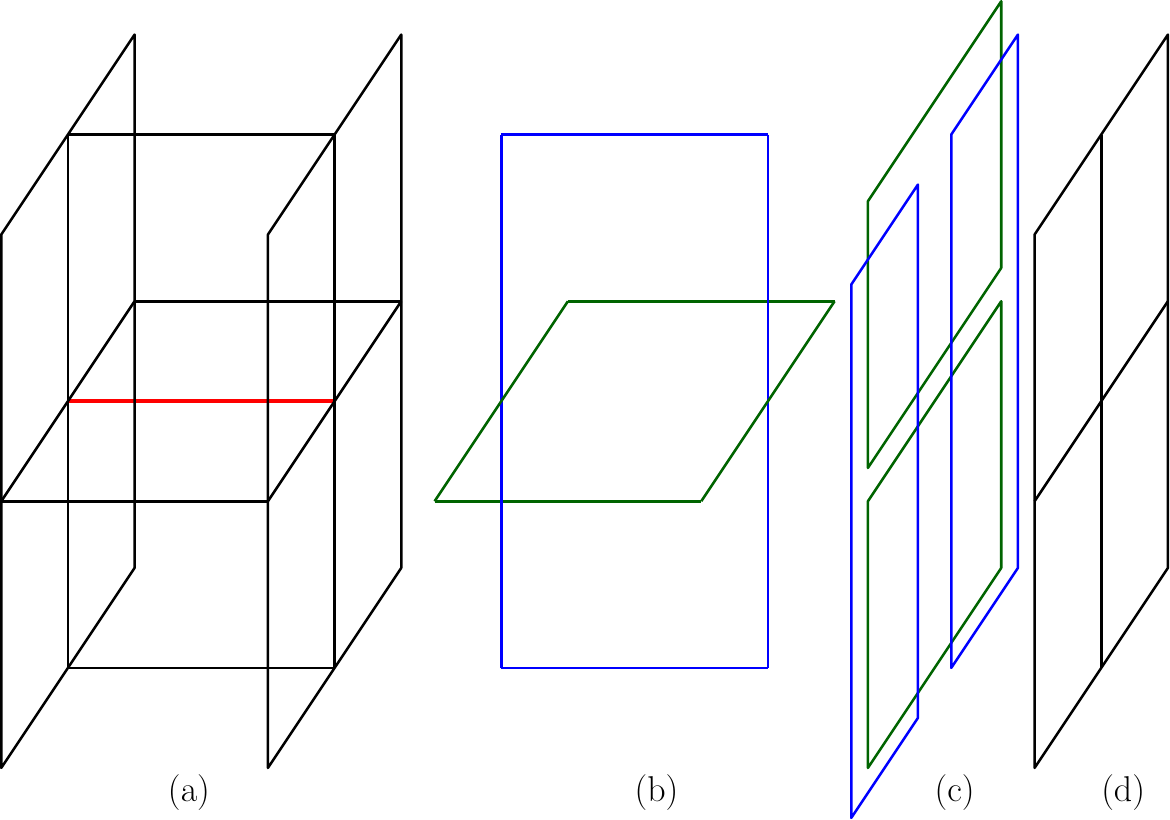}
	\caption{\label{fig:coefnet-input}Input to the neural network that computes smearing coefficients for updating the red link in a 3D lattice.
	From left to right, (a) the links in black used to compute Wilson loops as input to a neural network,
	(b) two 6-link rectangle loops parallel to the red link, (c) four 6-link rectangle loops perpendicular to the red link on one side,
	(d) four plaquette perpendicular to the red link on one side.}
\end{figure}

To limit implementation complexity,
we use only plaquette and $2\times 1$ rectangular traced Wilson loops for computing local smearing coefficients,
as shown in figure~\ref{fig:coefnet-input} for updating the red link (showing a 3D lattice as an example).
In a 4D space-time lattice gauge field,
for each link to be updated,
we use four plaquettes in each of the three planes parallel to and on each side of the updating link,
and the same amount of the $2\times 1$ rectangular loops in those planes,
and three $2\times 1$ rectangular loops that are parallel to the updating link.
These amount to 51 distinct loops that are locally close to the updating link.
To directly use available neural network implementations from TensorFlow~\cite{tensorflow2015-whitepaper},
we compute the traces of the ordered product, and the traces of their squared and cubic powers, of the gauge links along the loops,
and then use the real and imaginary parts of the complex numbers as the input to the traditional neural networks,
resulting in six real floating point numbers per Wilson loop, and a total of 306 real numbers per lattice site.
\begin{align}
	Y_{x,\mu,l} &= \{\re y_{x,\mu,l,i}, \im y_{x,\mu,l,i}\}_{i=1,2,3} \\
	y_i &= \tr W_{x,\mu,l}^i
\end{align}

Instead of directly calling 4D convolutional neural networks,
we implement symmetrical shift,
which shifts all the traces of Wilson loops of a fixed distance to the lattice link to be updated,
and then call dense neural networks,
\begin{equation}
	Y_{x,\text{out}} = \sigma(W Y_{x,\text{in}} + b),
\end{equation}
where the weight matrix $W$ and bias vector $b$ are shared among the lattice sites inside a particular layer.
We choose Swish~\cite{ramachandran2017searching} as the activation function, $\sigma$, for its continuous derivative.
We omit the index $\mu$ here, since we always update links in a single direction per neural network layer.

We also introduce a normalization layer similar to typical layer-normalization in computer vision,
but we use site-local scaling coefficients ($\gamma_i$) and shift ($\beta_i$),
which have the same dimension as the vector at each site and have the same values across the lattice.
\begin{align}
	m &= \frac{1}{DV/2} \sum_{x,i} Y_{x,i,\text{in}}, \\
	\tilde{Y}_{x,i} &= Y_{x,i,\text{in}} - m \\
	\sigma_m^2 &= \frac{1}{DV/2} \tilde{Y}_{x,i}^2 \\
	Y_{x,i,\text{out}} &= \gamma_i \tilde{Y}_{x,i}/\sigma_m + \beta_i
\end{align}
where $D$ is the vector dimension per site, $V$ is the lattice volume,
and the summation goes through the lattice (single parity) and the vector index.

Additionally, we implemented self-attention, feedforward, and residue networks,
which operate on numerical vectors on a 4D lattice.
However, neural network layers using these operations
exceed the available compute and memory
for reasonably large lattice volumes.

As we update the gauge links in a single direction on either even or odd lattice sites in a layer,
this computation creates a vector of real floating-point numbers on each lattice site of a given parity, even or odd.
Each operation---symmetric shift, site-local dense network, normalization, and etc.---applies to the lattice field of vectors on a single parity and returns a vector field of the same lattice size with optionally different degrees of freedom per site.
Thus, we can compose many layers of sequential operations that create the local coefficients used for gauge field transformation in the smearing update as in equation~\eqref{eq:transform}.
We sequentially apply transformation layers with different weights
to gauge link subsets in a single direction and parity.
This transforms the whole lattice gauge field,
yielding a continuously differentiable, bijective deep neural network map of the fields.

Our code for HMC with neural network field transformation is available online~\cite{nthmc}.

\section{Test result}

We use the DBW2 action with $c_1=-1.4088$, $\beta=0.7796$,
corresponding to $a/r_0\simeq 0.4$ and $a=0.2000(20)$ as a baseline following Ref~\cite{NECCO2004137,McGlynn:2014bxa}.
We optimize the map so that the gauge force of the effective action
matches an action with a stronger gauge coupling,
\begin{align}
	\label{eq:force-match}
	S'_{\text{FT}}(V;\beta=0.7796)
	&= \partial_{x,\mu} \left[S\big(\mathcal{F}(V;\beta=0.7796)\big) - \ln\big|\mathcal{F}_*(V)\big|\right] \\
	&\sim S'(V;\beta=\beta_{\text{T}}).
\end{align}
As $\beta$ is a multiplicative constant in the gauge action,
matching forces effectively seeks a transformation
that reduces the gauge force per link by a constant factor.

Denote the difference in force computed from the transformed action and the original action
with different couplings as,
\begin{equation}
	\label{eq:diff-force}
	\Delta_{x,\mu,c} = \partial_{x,\mu,c} S_{\text{FT}}(V;\beta) - \partial_{x,\mu,c} S(V;\beta_{\text{T}}),
\end{equation}
with the subscript $c$ denote the degree of freedom of the gauge Lie algebra,
and per link ($x,\mu$) l2-norm as,
\begin{equation}
	\label{eq:diff-force-norm}
	\Delta_{x,\mu} = \left(
	\sum_c \big(
		\partial_{x,\mu,c} S_{\text{FT}}(V;\beta) - \partial_{x,\mu,c} S(V;\beta_{\text{T}})
	\big)^2
	\right)^{1/2}.
\end{equation}
We choose the exponential of this l2-norm per link as the loss function for optimization:
\begin{equation}
	\label{eq:loss-lme-force-norm}
	L_{\text{LMEN}} =
	\log \sum_{x,\mu} \exp \big(\Delta_{x,\mu}\big)
	- \log (4 \text{Vol}).
\end{equation}

For training the models,
we generate configurations at $\beta=0.7796$, with a lattice size of $8^3\times 16$,
using 4 HMC streams, and a trajectory length of 4, saving configurations every 16 trajectories,
or 64 molecular dynamic time units,
which is larger than the integrated autocorrelation lengths reported in reference~\cite{McGlynn:2014bxa}.
The computed topological charges show negligible autocorrelation.
During training, we use the target $\beta_{\text{T}} = 0.7099$,
corresponding to $a/r_0\simeq 0.6$,
from an extrapolation formula in Ref~\cite{NECCO2004137}.

In a four-dimensional lattice, our code currently implement the transformation, using a stack of smearing layers, each using the derivatives of 6 plaquette and 48 chair terms for the smearing kernel per updating gauge link.
The coefficients can be globally tuned numbers ($6+48=54$ parameters for a single link) or the output of a constructed neural network as described in the previous section that depends on nearby traced Wilson loops.
In the following, we show two different models of gauge field transformation: one with global coefficients (GC) and one with a Dense-Normalization-Dense neural network (NN), each consisting of a stack of layers.

The order of layers in the GC model in Python's list comprehension is:
\begin{footnotesize}
\begin{verbatim}
transform=transform.TransformChain(
    [StoutSmearSlice(coeff=CoefficientVariable(p0, chair=c0, rng=rng), dir=dir, is_odd=eo)
        for _ in range(3) for dir in range(4) for _ in range(3) for eo in {False,True}]),
\end{verbatim}
\end{footnotesize}
The layers update the gauge link in a single direction for the even and then odd subset 3 times.
These are repeated for 4 different directions, covering the whole lattice gauge links.
Overall, the whole lattice is updated 3 times.
With 54 real parameters per layer, this transformation model uses $54\times 2\times 3\times 4\times 3 = 3888$ parameters.
The number of parameters is independent of the lattice size.

The order of layers in the NN model is:
\begin{footnotesize}
\begin{verbatim}
transform=transform.TransformChain(
    [StoutSmearSlice(coeff=CoefficientNets([Dense(units=8, activation='swish'),
                                            Normalization(),
                                            Dense(units=54, activation=None)]), dir=dir, is_odd=eo)
        for _ in range(4) for dir in range(4) for _ in range(1) for eo in {False,True}]
\end{verbatim}
\end{footnotesize}
with the coefficients generated by a Dense network, Normalization, and another Dense network.
The \verb|units| in the Dense network initialization denotes the dimension of the output vector.
The last Dense network does not use a non-linear activation function,
as its 54 output numbers directly feed into the input of the $\tan^{-1}$ function in equation~\eqref{eq:def-eps}.

We train our models using a single Nvidia A100 GPU with 40 GB memory.
Training the GC model uses 64 gauge configurations for 8 epochs, with each epoch going through the 64 configurations in a random order.
It takes 3 hours and uses a maximum of 30 GB GPU memory.
The NN model, on the other hand, has more parameters and uses 256 configurations for training over 16 epochs.
That takes 11 hours and uses a maximum of 20 GB GPU memory.

During evaluation on a single Nvidia A100 GPU with 40 GB memory, we use gauge configurations generated with HMC at $\beta=0.7796$ for $8^3\times 16$ lattices, and $\beta=0.8895$ for $12^3\times 24$ lattices, using the DBW2 gauge action, from separate HMC streams that are different from those used for training.
With a lattice size of $8^3\times 16$,
a single force evaluation of the effective action, equation~\eqref{eq:effective-action}, with the GC model takes 1.6 seconds and uses a maximum of 10 GB GPU memory.
A single force evaluation with the NN model takes 1.1 seconds and uses a maximum of 8 GB GPU memory.
To compare, using the same framework,
a single force evaluation of the DBW2 lattice action without transformations takes less than 0.005 seconds and uses a maximum of 128 MB GPU memory.
The evaluation of the same models on the $12^3\times 24$ lattices requires more memory than the 40 GB provided by the single Nvidia A100 we use,
and thus runs on CPUs.

Since our machine learning framework (TensorFlow) has limited ability to distribute lattice gauge configurations across GPUs,
and our models have no lattice size dependencies,
we focus on how the machine-learned model scales with lattice size.

\begin{figure}
	\centering
	\begin{subfigure}{.49\linewidth}\centering\includegraphics[width=\linewidth]{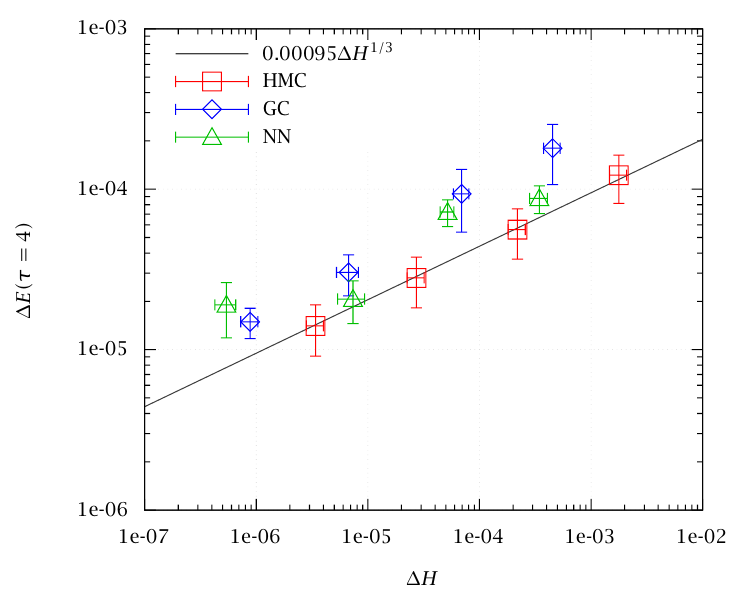}
		\caption{$8^3\times 16$}
	\end{subfigure}
	\begin{subfigure}{.49\linewidth}\centering\includegraphics[width=\linewidth]{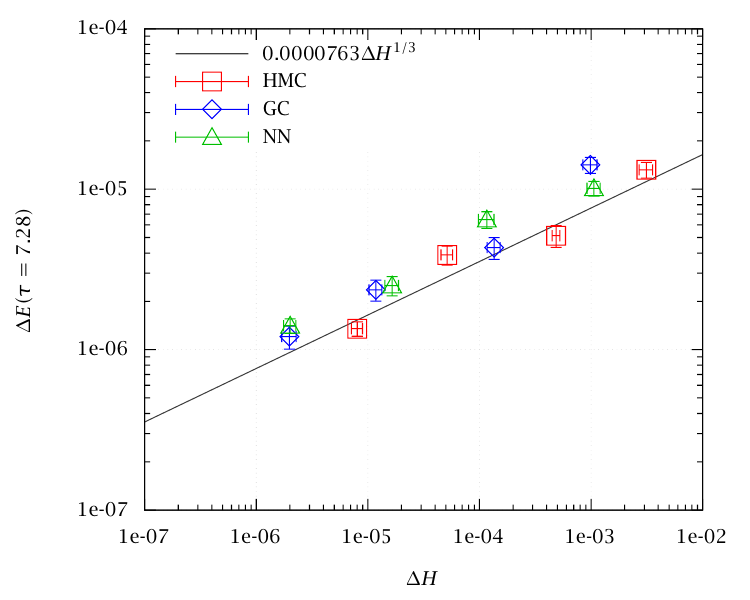}
		\caption{$12^3\times 24$}
	\end{subfigure}
	\caption{\label{fig:FT-dH_dE} The difference in energy density versus the difference in Hamiltonian after a single Omelyan step.  The energy density is at Wilson flow time, $\tau = 4$ for $8^3\times 16$ with $\beta=0.7796$, and $\tau = 7.28$ for $12^3\times 24$ with $\beta=0.8895$.  The step sizes for the Omelyan 2MN integrator are $\delta t = 0.0025$, $0.005$, $0.01$, and $0.02$, for symbols from left to right.}
\end{figure}

To study the effects of field transformation on the molecular dynamics in HMC, we start from thermalized configurations and perform a single step of the Omelyan 2MN integrator~\cite{OMELYAN2003272}.
We measure the energy density using the lattice field tensor computed at 1-loop order with four plaquette loops~\cite{Luscher:2010iy} at Wilson flow time, $\tau = 4$, and examine how the integration step changes the energy density and the Hamiltonian.
Figure~\ref{fig:FT-dH_dE} shows the difference in energy density versus the difference in Hamiltonian, before and after the integration step.
The three kinds of symbols are from the lattice action without transformation (labeled ``HMC''),
the lattice effective action with the field transformation with global coefficients (labeled ``GC''),
and the lattice effective action with the field transformation with a Dense-Normalization-Dense neural network (labeled ``NN'').
The same kind of symbols from left to right correspond to 
four different step sizes, $\delta t = 0.0025$, $0.005$, $0.01$, and $0.02$.
With a single integration step,
$\Delta H$ scales with $\delta t^3$, and $\Delta E$ scales with $\delta t$, thus $\Delta E \propto \Delta H^{1/3}$.
We fit the line through the HMC data and show it in the figure.

The left panel in figure~\ref{fig:FT-dH_dE} shows the model performance on $8^3\times 16$ at $\beta=0.7796$, while the right shows performance on $12^3\times 24$ at $\beta=0.8895$,
which has $a/r_0\simeq 0.265$ and $a=0.1326(13)$ following Ref~\cite{NECCO2004137,McGlynn:2014bxa},
roughly keeping the same physical volume.
The trained model achieves smaller $\Delta E$ and $\Delta H$ with the same step size.
For $8^3\times 16$ lattices, the models manage to gain larger $\Delta E$ at much smaller $\Delta H$, indicating a potential factor of a few better than HMC without field transformations,
keeping $\Delta H$ constant while inducing a larger change in Wilson-flowed energy density.
In addition, the same models trained on $8^3\times 16$ work well with the $12^3\times 24$ lattices,
maintaining a decent speedup with the larger lattice volume.

\section{Conclusion}

We implement and optimize gauge field transformation models for reduced effective gauge forces.
Applying the machine-optimized models as a change of variables in HMC
reduces the Hamiltonian violation, $\Delta H$,
and increases the difference in Wilson-flowed energy, $\Delta E$, in our preliminary results.

The improvements remain when we directly apply the models to lattice configurations larger than the lattice volume used for training,
indicating the transferability of the learned transformations.
With further improvements in evaluation metrics, frameworks, and transformation models,
this approach remains a valuable research direction for enhancing sampling efficiency in lattice QCD simulations.

\acknowledgments

We thank
Peter Boyle, Norman Christ, Sam Foreman, Taku Izubuchi,
Luchang Jin, Chulwoo Jung, James Osborn, Akio Tomiya,
and other ECP collaborators for insightful discussions and support.
This research was supported by the Exascale Computing Project
(17-SC-20-SC), a collaborative effort of the U.S. Department of
Energy Office of Science and the National Nuclear Security
Administration.
Part of this work was done on a pre-production supercomputer with early versions of the Aurora software development kit.
This research used resources of the Argonne Leadership Computing
Facility, a U.S. Department of Energy (DOE) Office of Science user
facility at Argonne National Laboratory and is based on research
supported by the U.S. DOE Office of Science-Advanced Scientific
Computing Research Program, under Contract No. DE-AC02-06CH11357.
We gratefully acknowledge the computing resources provided and operated by the Joint Laboratory for System Evaluation (JLSE) at Argonne National Laboratory.

\bibliographystyle{JHEP}
\bibliography{ref}

\end{document}